\documentclass[12pt]{article}
\usepackage{latexsym}
\usepackage{amsmath,,calrsfs}
\usepackage{amsfonts}
\usepackage{amssymb}
\usepackage{amscd}
\usepackage{bbm}
\usepackage{fancybox}
\usepackage{cite}
\usepackage{amsmath,amsfonts,amsbsy}
\usepackage{pstricks,pst-node}
\usepackage[small,bf,hang]{caption2}
\usepackage{graphicx}
\usepackage{epsfig}
\usepackage{psfrag}
\usepackage{comment}

\usepackage{float}

\psset{unit=1.3cm,linewidth=.5pt,radius=.2}  

\usepackage{multirow}                     
\usepackage{float}                          
\usepackage{lscape}                         
\usepackage{bm}

\newcommand{\bb}{\bar\beta}

\newcommand{\beq}{\begin{equation}}
\newcommand{\eeq}{\end{equation}}
\newcommand{\bi}{\begin{itemize}}
\newcommand{\ei}{\end{itemize}}
\newcommand{\bt}{\begin{tabular}}
\newcommand{\et}{\end{tabular}}
\newcommand{\bc}{\begin{center}}
\newcommand{\ec}{\end{center}}

\newcommand{\be}{\begin{equation}}
\newcommand{\ee}{\end{equation}}
\newcommand{\bea}{\begin{eqnarray}}
\newcommand{\eea}{\end{eqnarray}}
\newcommand{\ba}{\begin{array}}
\newcommand{\ea}{\end{array}}

\def\bbox{{\,\lower0.9pt\vbox{\hrule \hbox{\vrule height 0.2 cm
\hskip 0.2 cm \vrule height 0.2 cm}\hrule}\,}}
\newcommand{\dsl}{\pa \kern-0.5em /}

\font\mybb=msbm10 at 12pt
\def\bb#1{\hbox{\mybb#1}}
\def\bZ {\bb{Z}}
\def\bR {\bb{R}}

\def\bP {\bb{P}}

\def\bI{\bb{I}}

\newcommand{\Pslash}{P\hskip -.24truecm  / }
\newcommand{\Xslash}{X\hskip -.24truecm  / }
\newcommand{\calPslash}{{\cal P}\hskip -.29truecm  / }

\makeatletter \@addtoreset{equation}{section} \makeatother

\def\slashchar#1{\setbox0=\hbox{$#1$}           
   \dimen0=\wd0                                 
   \setbox1=\hbox{/} \dimen1=\wd1               
   \ifdim\dimen0>\dimen1                        
      \rlap{\hbox to \dimen0{\hfil/\hfil}}      
      #1                                        
   \else                                        
      \rlap{\hbox to \dimen1{\hfil$#1$\hfil}}   
      /                                         
   \fi}

\begin{document}

\begin{titlepage}
\begin{center}

\hfill  DAMTP-2016-46

\vskip 1.5cm

{\Large \bf  Worldline CPT and massless supermultiplets}

\vskip 2cm

{\bf Alex S. Arvanitakis\,${}^1$, Luca Mezincescu\,${}^2$,  and 
Paul K.~Townsend\,${}^1$} \\

\vskip 15pt

{\em $^1$ \hskip -.1truecm
\em  Department of Applied Mathematics and Theoretical Physics,\\ Centre for Mathematical Sciences, University of Cambridge,\\
Wilberforce Road, Cambridge, CB3 0WA, U.K.\vskip 5pt }

{email: {\tt A.S.Arvanitakis@damtp.cam.ac.uk, P.K.Townsend@damtp.cam.ac.uk}} \\

\vskip .4truecm

{\em $^2$ \hskip -.1truecm
\em Department of Physics
University of Miami, \\ Coral Gables, FL 33124, USA\vskip 5pt }

{email: {\tt Mezincescu@physics.miami.edu}} \\

\end{center}

\vskip 0.5cm

\begin{center} {\bf ABSTRACT}\\[3ex]
\end{center}

The action for a massless particle in 4D Minkowski spacetime  has a worldline time-reversing symmetry corresponding to  CPT invariance of  the quantum theory.
The analogous symmetry of the  ${\cal N}$-extended superparticle is shown to be anomalous when ${\cal N}$ is odd; in the supertwistor formalism this is because 
a CPT-violating worldline-Chern-Simons  term is needed to preserve the chiral $U(1)$ gauge invariance. This accords with the fact that no massless ${\cal N}=1$  
super-Poincar\'e irrep is CPT-self-conjugate.  There {\it is}  a CPT self-conjugate supermultiplet when ${\cal N}$ is even, but it has $2^{{\cal N}+1}$ states when 
$\tfrac{1}{2}{\cal N}$ is odd (e.g. the ${\cal N}=2$ hypermultiplet)  in contrast to just $2^{\cal N}$ when $\tfrac{1}{2}{\cal N}$ is even (e.g. the ${\cal N}=4$ Maxwell 
supermultiplet). This is shown to follow from a Kramers degeneracy of the  superparticle state space when $\tfrac{1}{2}{\cal N}$ is odd.

\end{titlepage}

\newpage
\setcounter{page}{1} 
\tableofcontents


\section{Introduction}

In a first course on supersymmetry one learns how to construct supermultiplets of one-particle states  directly from the ${\cal N}$-extended supersymmetry algebra. For a massless particle in a four-dimensional Minkowski  spacetime with given null $4$-momentum, each of the ${\cal N}$ spinor charges has two independent components, with linear combinations that can be interpreted as operators that raise or lower helicity by $\frac{1}{2}$ (in units where $\hbar=1$). The ${\cal N}$ helicity lowering operators  are mutually anticommuting and annihilate a ``Clifford vacuum'' of some definite helicity $h$. By acting on this state with the helicity raising operators,  one constructs an irreducible supermultiplet of $2^{\cal N}$ independent states with helicities ranging from $h$ to $h+ \tfrac{1}{2} {\cal N}$.  

For example, if ${\cal N}=4$ and we choose $h=-1$ then this construction generates the CPT self-conjugate ${\cal N}=4$ Maxwell supermultiplet with $16$ independent helicity states, of which two are the
polarisation states of a massless spin-$1$ particle.  If ${\cal N}=2$ the choice $h=\tfrac{1}{2}$ yields, apparently, a CPT self-conjugate ${\cal N}=2$ supermultiplet with four states of helcities 
$(-\tfrac{1}{2},0,0,\tfrac{1}{2})$. However, the only CPT self-conjugate massless ${\cal N}=2$ supermultiplet is the hypermultiplet, which is a ``doubled'' version of the 
$(-\tfrac{1}{2},0,0,\tfrac{1}{2})$ supermultiplet with four spin-$\tfrac{1}{2}$ states and four spin-$0$ states.  The reason for this is that an irreducible  CPT self-conjugate  supermultiplet  actually has 
only $2^{\cal N}$ independent helicity states if it satisfies a reality condition, and this is possible for ${\cal N}=4$ but not for ${\cal N}=2$.  

For ${\cal N}=1$ the construction described above yields a multiplet of helicities $(h,h+\tfrac{1}{2})$, which is  the irreducible supermultiplet of   ``superhelicity'' $h$ \cite{Buchbinder:1998qv}.
As $2h$ must be an integer, no such multiplet can be CPT self-conjugate. It follows that any  CPT self-conjugate ${\cal N}=1$ supermultiplet is {\it necessarily} reducible.  For example,  the massless Wess-Zumino model \cite{Wess:1973kz} has ${\cal N}=1$ supersymmetry realised on  fields whose quantum 1-particle states belong to a supermultiplet of helicities $(-\tfrac{1}{2}, 0,0, \tfrac{1}{2})$, which is the direct sum of irreducible ${\cal N}=1$ supermultiplets of superhelicities $0$ and $-\tfrac{1}{2}$. 

The aim of this paper is to show how these peculiarities of the representation theory of ${\cal N}$-extended supersymmetry follow directly from properties of the  ${\cal N}$-extended massless superparticle. 
The classical action is invariant, for any ${\cal N}$, under a composition of  worldline time reversal with space-and-time  reversal (PT); this becomes CPT in the quantum theory because  
worldline time reversal is realised quantum mechanically via an {\it anti-unitary} operator. However, we shall see that 
there is a global CPT anomaly for odd ${\cal N}$, which explains why there is no CPT  self-conjugate ${\cal N}=1$ supermultiplet. 

The CPT anomaly is most clearly seen in the supertwistor formulation of superparticle mechanics \cite{Shirafuji:1983zd}. In this  formulation  there is a chiral $U(1)$ gauge invariance associated to a
phase-space ``spin-shell'' constraint that specifies the superhelicity.  One can add to this action a CPT-violating Worldline-Chern-Simons (WCS) term, although  $U(1)$ gauge invariance forces its coefficient to be  an integer if ${\cal N}$ is even,  and an integer plus $\tfrac{1}{2}$ if ${\cal N}$ is odd.  It is therefore possible to omit the  WCS term, and hence preserve CPT, 
when ${\cal N}$ is even but not when ${\cal N}$ is odd; in the latter case, the clash between $U(1)$ gauge invariance and CPT is a close analogy  of  the  clash between Yang-Mills gauge invariance and parity in 3D \cite{Redlich:1983kn}. 

The CPT anomaly of superparticle mechanics may be cancelled, at the expense of irreducibility,  in much the same way that the parity anomaly of 3D Chern-Simons (CS)  theory may be cancelled: by arranging to have a sum of two  CS terms with equal but opposite sign coefficients \cite{Hagen:1991ku}.  In our case we must add two superparticle actions  with WCS coefficients that 
sum to zero. Quantization then yields a reducible but CPT self-conjugate supermultiplet of the type required for a field theory that is local in Minkowski spacetime. 

Although worldline-CPT may be preserved for even ${\cal N}$ without sacrificing irreducibility, its realization via an anti-unitary operator $K$ has implications. On general grounds, $K^2=\pm1$ and if $K^2=1$ it is possible to impose the condition $K|\Psi\rangle= |\Psi\rangle$ on the superparticle wavefunction $|\Psi\rangle$, but this is not possible if $K^2=-1$; in such cases  there is a doublet degeneracy called ``Kramers degeneracy''.  We show here that 
\begin{equation}
K^2= (-1)^{\tfrac{{\cal N}}{2}} \qquad \left(\tfrac{1}{2} {\cal N} \in \bZ^+\right)\, , 
\end{equation}
which means that $K^2=1$ for ${\cal N}=4$ but $K^2=-1$ for ${\cal N}=2$. Consequently there is a Kramers degeneracy for ${\cal N}=2$, which explains the doubling of states needed for the 
hypermultiplet. 

Our detailed presentation of these results is prefaced, in the following section,  by a discussion of worldline time-reversal  for the bosonic and spinning particle in which we aim to clarify 
issues that arise when anticommuting variables are involved.  Then, in the subsequent two sections we review relevant features of the massless 4D superparticle action, and  present its supertwistor formulation in a Majorana spinor notation that simplifies the verification of `` worldline-CPT'' invariance. This informs our discussion of the quantum theory in section 5, where we articulate how the clash between the $U(1)$ gauge invariance and worldline-CPT leads to CPT violation for odd ${\cal N}$. We also discuss how the anomaly may be cancelled for ${\cal N}=1$, and 
how  Kramers degeneracy leads to ``doubled''  CPT-conjugate supermultiplets  for odd $\frac{1}{2}{\cal N}$, in particular ${\cal N}=2$. 
We end with a brief summary of our results and a discussion of some open questions.

\section{Time reversal preliminaries}\label{sec:prelim}

Let us begin with the zero-spin massless particle in a 4-dimensional Minkowski  spacetime with metric $\eta_{mn}$ for cartesian coordinates $\{X^m;m=0, 1,2,3\}$.  Its worldline-time reparametrization invariant phase-space action for arbitrary parameter  interval $2T$ is 
\begin{equation}\label{bosact}
S= \int_{- T}^T dt \left\{ \dot X^m P_m - \frac{1}{2} e \, P^2\right\} \, , 
\end{equation}
where $e(t)$ is a Lagrange multiplier for the mass-shell constraint $\eta^{mn}P_mP_n=0$. This action is invariant under the composition of
worldline-time reversal and spacetime inversion:
\begin{equation}\label{CPT1}
t\to -t\, , \quad e(t)\to e(-t) \, ; \quad X(t) \to  -X(-t)\, , \quad P(t) \to P(-t)\, .
\end{equation}
Notice that the integral $\int_{-T}^T dt$ is invariant (and equal to $2T$) because the change in the integration limits produces a sign that cancels the sign change of $dt$.  The ``geometric''  ($\dot X P)$ term in the Lagrangian would change sign because of the derivative with respect to the worldline time parameter  if it were not 
for the additional change in sign of $X$. 

In the quantum theory this symmetry transformation  is realised by conjugation with an anti-unitary operator $K$ (see e.g. \cite{Sachs:1987gp}). 
In a basis for which the momentum 
operator $\hat P$ is diagonal we have  $K=K_0$, where $K_0$  is the operation of complex conjugation.  In this basis the physical state condition 
$\hat P^2|\Psi\rangle =0$ becomes the mass-shell constraint $P^2=0$, and $K$ takes the momentum space wavefunction 
$\Psi(P) \equiv \langle P|\Psi\rangle$ to its complex conjugate.  In general $K^2=\pm1$ but in this case $K^2=1$, so we may impose
the condition $K|\Psi\rangle = |\Psi\rangle$ which implies that $\Psi(P)$ is real. 

Notice that the action (\ref{bosact}) has an independent ``internal'' symmetry that flips the sign of both $X$ and $P$. If we compose this with the symmetry transformation
(\ref{CPT1}) then we arrive at the transformations
\begin{equation}\label{CPT2}
t\to -t\, , \quad e(t)\to e(-t) \, ; \quad X(t) \to  X(-t)\, , \quad P(t) \to -P(-t)\, .
\end{equation}
This is again realised by an anti-unitary operator $K$ in the quantum theory but  we now have $K=K_0$ in a basis for which the operator $\hat X$ is diagonal, and the condition 
$K|\Psi\rangle = |\Psi\rangle$ is now equivalent to reality of the wavefunction $\tilde\Psi(X)= \langle X| \Psi\rangle$,  the Fourier transform of $\Psi(P)$. 
The transformation (\ref{CPT2})  is the worldline time reversal transformation used in \cite{Henty:1988hh} for the analysis of the quantum mechanics of the worldline-supersymmetric ``spinning particle'', which we shall revisit below. Here we prefer to use  (\ref{CPT1}), mainly  because it  simplifies our analysis of the massless superparticle, but also because the lightcone gauge-fixing condition $X^+(t)=t$ is invariant  under the transformations of (\ref{CPT1}) but not those of (\ref{CPT2}). 

We shall be considering actions that generalize (\ref{bosact}) via the addition of terms involving anticommuting variables, so let us first consider 
such ``fermionic'' contributions  in isolation.  In particular, consider the following action depending on $2n$ ``real'' anticommuting variables $\{\psi^I; I=1,2, \dots 2n\}$
for some positive integer $n$:
\begin{equation}\label{fermaction}
S[\{\psi\}] = \int dt \left\{ i \delta_{IJ} \psi^I\dot\psi^J  -H_F\right\} \,  \qquad H_F= i m_{IJ}\psi^I\psi^J +g_{IJKL}  \psi^I\psi^J\psi^K\psi^L\, , 
\end{equation}
for (totally antisymmetric) real constants $m_{IJ}$ and $g_{IJKL}$,  where a sum over repeated indices is implicit. The factors of $i$ are needed because of  the standard convention that complex conjugation of a product of anticommuting variables reverses their  order. In the quantum theory we have
\begin{equation}
\hat H_F = i m_{IJ}\hat\psi^I\hat\psi^J+ g_{IJKL}\hat\psi^I\hat\psi^J\hat\psi^K\hat\psi^L\, , 
\end{equation}
where the operators $\hat\psi^I$ satisfy the canonical anti commutation relations
\begin{equation}\label{cacrs}
\left\{\hat\psi^I,\hat\psi^J\right\} = \delta^{IJ}\, . 
\end{equation}

Let us first consider the effect of a $t$-reversing symmetry on the quantum theory, where it will be realised by an anti-unitary operator $K$. We will suppose that 
\begin{equation}\label{conjugation}
K\hat\psi^I K^{-1} = \sigma_{(I)} \hat \psi^I\qquad I=1,\dots,2n\, , 
\end{equation}
where $\sigma_{(I)}$ are signs.  Notice that this transformation preserves the canonical anti-commutation relations (\ref{cacrs}).  Taking into account that $KiK^{-1}=-i$, we see that $K$ commutes 
with $\hat H_F$  iff
\begin{equation}\label{Witten's}
\sigma_{(I)} m_{IJ} = \sigma_{(J)}m_{JI} \qquad \&\qquad \left[\sigma_{(I)} \sigma_{(J)}\sigma_{(K)} \sigma_{(L)}\right]g_{IJKL}= g_{IJKL}\, . 
\end{equation}
This tells us that the signs $\sigma_{(I)}$ are arbitrary if $\hat H_F=0$, but otherwise they are restricted. In particular, each non-zero 
skew-eigenvalue  of the antisymmetric matrix $m_{IJ}$ must  correspond  to a pair of anticommuting variables $(\psi,\psi')$ such that 
$\psi'$ transforms with the opposite  sign to $\psi$, in agreement with \cite{Witten:2015aba}. 

To summarise, the criterion for time reversal invariance in a quantum theory with fermionic variables is essentially the same  as it is for any other quantum theory:  the quantum Hamiltonian should commute 
with the anti-unitary operator $K$ representing time-reversal and the canonical (anti)commutation relations must be invariant under conjugation of the canonical variables with $K$.  
However, a special feature of ``fermionic'' models emerges when one asks the following question: what  transformation of the anticommuting variables $\psi^I$ in the classical action (\ref{fermaction}) corresponds to conjugation of the operators $\hat\psi^I$  by the anti-unitary operator $K$?

One might suppose that the classical analog of  (\ref{conjugation}) should be $\psi^I(t) \to \sigma_{(I)} \psi^I(-t)$, but this leads to a change in sign of the ``geometric''  term in the action
(\ref{fermaction}),  and hence to a change in sign of the anticommutation relations. It also leads to a change in sign of any mass term with matrix 
$m_{IJ}$ satisfying the first of the conditions of  (\ref{Witten's}).  For $g_{IJKL}=0$ this would imply a change in sign of $H_F$ and hence of the entire action, which 
would at least imply invariance  of the equations of motion, but even this fails for  non-zero $g_{IJKL}$ because  the  quartic term in anticommuting variables is invariant; it does {\it not} change sign.  

A more general way to understand the nature of the problem is by consideration of Dirac's rule for canonical quantization, usually expressed in the schematic form
\begin{equation}
\left\{ , \right\}_{PB} \to -i \left[ , \right]\, . 
\end{equation}
If we apply this rule to classically-commuting observables that are either even or odd under time reversal then the right hand side changes by a product of minus signs under 
time reversal with one  minus sign arising from the fact that $KiK^{-1} =-i$. On the left hand side we get this extra minus sign from the definition of the Poisson bracket in 
terms of the symplectic 2-form 
defined by the classical phase-space action; it is the same minus sign that is needed for invariance of the geometric term in the action.  

Let us now attempt to apply the same reasoning  to the action (\ref{fermaction}). In this case we use the ``fermionic Dirac rule'' 
\begin{equation}\label{Dirac2}
\left\{ , \right\}_{PB} \to -i \left\{ , \right\}\, ,
\end{equation}
where the Poisson bracket is now symmetric, being defined in terms of the invertible 2-form $i\delta_{IJ}d\psi^I \wedge d\psi^J$; because of the factor of $i$ in this expression, an application 
of the fermionic Dirac rule leads to the canonical anticommutation relations (\ref{cacrs}). Applying the  rule to classically-anticommuting observables 
that are either even or odd under time reversal we again get a transformation of the right hand side by  a product of minus signs, again with one sign arising from the fact that 
$KiK^{-1} =-i$. In contrast, a  classical time-reversal transformation of the form $\psi^I(t) \to \sigma_{(I)} \psi^I(-t)$ yields no corresponding minus sign on the left hand side
because it leaves invariant the 2-form $i\delta_{IJ}d\psi^I \wedge d\psi^J$ rather than changing its sign. 

But what alternative transformation is there that avoids these problems? The answer is simple. We must include a factor of $i$ in the transformation of each ``real''  anticommuting variable.  In other words, 
\begin{equation}
\label{fermi_transformation}
t\to -t \, ; \quad \psi^I(t) \to i\sigma_{(I)}\psi^I(-t)\, . 
\end{equation}
This leaves invariant all terms in the action (\ref{fermaction}), given (\ref{Witten's}). It also ensures that the ``fermionic'' Poisson bracket 
is intrinsically odd under time-reversal, as required by the fermionic Dirac rule.  Another effect of the factor of $i$ is that the transformation is no longer a $\bZ_2$ transformation 
because iteration  yields minus the identity, but this is not a problem because $\psi^I\to -\psi^I$ is also a symmetry\footnote{This may be a classical vestige of the
fact that the operator $K$ realizing time-reversal in the quantum theory must satisfy $K^4=1$ but need not satisfy $K^2=1$.}.

Of course,  it  would be inconsistent to transform a real  variable such as $x(t)$ into  an imaginary variable but when we say that an {\it anticommuting}
variable $\psi(t)$ is ``real''  we are merely stating a rule to be applied when taking the complex  conjugate of any equation involving $\psi$: this rule is $\psi^*=\psi$. However, we could equally well choose  the rule $\psi^*=-\psi$ because the action does not distinguish between these two possibilities.  The  classical worldline time reversal transformation of (\ref{fermi_transformation}) is also suggested by an observation  of \cite{Henty:1988hh}: if the quantum states are elements 
of a vector space over a supernumber field, then we require $K\lambda K^{-1}= i\lambda$ for ``real'' anticommuting number $\lambda$ in order that  $K$ commute with 
all ``real''  nilpotent supernumbers.

Let us now see how this resolution of the puzzle of classical time-reversal invariance for ``fermionic'' actions applies to the massless ``spinning particle'' 
in a $D$-dimensional Minkowski spacetime \cite{Brink:1976sz}.  Instead of the action (\ref{bosact}) we now have
\begin{equation}\label{bosact2}
S= \int_{- T}^T dt \left\{ \dot X^m P_m  + \frac{i}{2} \eta_{mn} \lambda^m  \dot\lambda^n  - \frac{1}{2} e \, P^2 - i\chi \lambda^m P_m \right\} \, ,  
\end{equation}
where $\lambda^m$ is an anticommuting $D$-vector variable and  $\chi$ is an anticommuting Lagrange multiplier for the Grassmann-odd constraint.  This action is invariant under the transformation\footnote{The transformation used  in \cite{Henty:1988hh} is the composition of this with the discrete symmetry that flips the sign of both $X$ and $P$.}
\begin{eqnarray}\label{classsym}
t\to -t\, ; &&\quad e(t)\to e(-t) \quad X(t) \to  -X(-t)\, , \quad P(t) \to P(-t) \nonumber \\ 
&&\quad  \lambda^m(t) \to  i \eta \lambda^m (-t) \qquad (\eta=\pm1)\, . 
\end{eqnarray}
The sign  $\eta$ must be the same for all components of the $D$-vector $\lambda$ in order for the transformations to be Lorentz covariant\footnote{The notation for this sign is chosen to agree with  
\cite{Kugo:1982bn}.}. 
In the quantum theory, the operators $\hat \lambda^m$ (hermitian for $m\ne0$ but anti-hermitian for $m=0$)
satisfy the canonical anticommutation relations 
\begin{equation}
\left\{ \hat \lambda^m,\hat\lambda^n\right\} = \eta^{mn}\, , 
\end{equation}
implying a spinorial wavefunction satisfying the massless Dirac equation \cite{Brink:1976uf}. 
The classical symmetry (\ref{classsym}) is realized in the quantum theory via an anti-unitary operator 
$K$ such that
\begin{equation}\label{Klam}
K\hat\lambda^m K^{-1} = \eta \hat\lambda^m\, . 
\end{equation}

Now we suppose that $D=2n+2$ and we choose the following new basis: 
\begin{equation}
\hat\lambda^\pm = \frac{1}{\sqrt{2}}\left(\hat\lambda^{2n+1}\pm \hat\lambda^0\right)\, , \qquad \hat\xi_i = \frac{1}{\sqrt{2}}\left(\hat\lambda_i + i \hat\lambda_{i+n}\right)\qquad (i=1,\dots,n).
\end{equation}
Notice that since $KiK^{-1}=-i$, it follows from  (\ref{Klam}) that 
\begin{equation}\label{conjugateK}
K\hat\xi_i K^{-1} = \eta \hat\xi_i^\dagger \, , \qquad K\hat\xi_i^\dagger  K^{-1} = \eta \hat\xi_i\, . 
\end{equation}
This transformation preserves the new canonical anticommutation relations 
\begin{equation}\label{anticomm}
\left\{ \hat\lambda^+,\hat\lambda^-\right\} =1\, , \qquad \left\{\hat\xi_i, \hat\xi_i^\dagger\right\} = 1 \quad (i=1,\dots,n)\, , 
\end{equation}
which can be realised by the $2^{n+1}\times 2^{n+1}$ matrices 
\begin{eqnarray}
\hat\lambda^\pm &=& \tfrac{1}{2} \sigma_\pm \otimes \bI_2 \otimes \bI_2 \otimes \cdots \otimes \bI_2 \otimes \bI_2 \nonumber \\
\hat\xi_1 &=&  \tfrac{1}{2} \sigma_3 \otimes \sigma_+ \otimes \bI_2 \otimes \cdots \otimes \bI_2 \otimes \bI_2 \nonumber \\
\hat\xi_2 &=& \tfrac{1}{2} \sigma_3 \otimes \sigma_3 \otimes \sigma_+ \otimes \cdots \otimes \bI_2 \otimes \bI_2 \nonumber \\
\vdots &=& \vdots \nonumber \\
\hat\xi_n &=&  \tfrac{1}{2} \sigma_3 \otimes \sigma_3 \otimes \sigma_3 \otimes \cdots \otimes \sigma_3 \otimes \sigma_+\, , 
\end{eqnarray}
where $\sigma_\pm=\sigma_1 \pm i \sigma_2$ and $\{\sigma_i; i=1,2,3\}$ are the Pauli matrices. For this  representation, the relations (\ref{conjugateK}) are satisfied 
by anti-unitary operator $K= B\circ K_0$, for unitary $2^{n+1}\times 2^{n+1}$ matrix $B$ of  the form 
\begin{equation}
B =   L^{\tfrac{1}{2}(1-\eta)}K_1 \cdots K_n\, , 
\end{equation}
where $L$ and $K_i$ ($i=1,\dots,n)$ are the {\it real symmetric} matrices 
\begin{eqnarray}
L &=& \sigma_3 \otimes\sigma_3 \otimes \sigma_3 \otimes \cdots \otimes \sigma_3\otimes \sigma_3\,  \nonumber \\
K_1 &=& \bI_2 \otimes \sigma_1 \otimes \bI_2 \otimes \cdots \otimes \bI_2\otimes \bI_2\nonumber \\
K_2 &=& \bI_2 \otimes \sigma_3 \otimes \sigma_1 \otimes \cdots \otimes \bI_2\otimes \bI_2\nonumber \\
\vdots &=& \vdots \nonumber \\
K_n &=&  \bI_2 \otimes \sigma_3 \otimes \sigma_3 \otimes \cdots \otimes \sigma_3 \otimes \sigma_1\, . 
\end{eqnarray}
Notice that  
\begin{equation}
L^2= \bI\, , \quad \{ L,K_i\} =0\, , \quad \left\{K_i,K_j\right\}=\delta_{ij} \, , 
\end{equation}
which implies (since $B$ is real) that 
\begin{equation}
K^2 = \eta^n(-1)^{\tfrac{n(n-1)}{2}}\, . 
\end{equation}
This tells us  that $K^2=1$ for $n=1$,  if we choose $\eta=1$,  and for $n=4$ irrespective of the choice for $\eta$. It also tells us that  $K^2=-1$ for $n=2$,  again irrespective of the choice for $\eta$; as observed in  \cite{Henty:1988hh}, this result is equivalent to the well-known fact  that  Majorana spinors do not exist for $D=6$. We will later use a variant of this argument in our analysis of the massless superparticle.

\section{The massless superparticle}\label{sec:sparticle}

The action of the  4D massless superparticle with ${\cal N}$-extended supersymmetry is \cite{Casalbuoni:1976tz,Brink:1981nb}
\be\label{CBS}
S= \int dt \left\{\left(\dot X^m - i \bar\Theta_i\Gamma^m \dot\Theta_i\right)P_m  - \frac{1}{2} e\, P^2\right\}\, , 
\ee
where $\Theta_i(t)$ ($i=1, \dots, {\cal N}$) are anticommuting Majorana spinors (summation over repeated indices is implicit) and $\bar\Theta_i=\Theta_i^TC$, for charge conjugation matrix 
$C$. In a  real representation of the 4D Dirac matrices $\Gamma^m$ we may choose\footnote{This is an opposite sign convention for a Majorana conjugate spinor as compared to \cite{Mezincescu:2013nta}, 
which accounts for a sign difference in the action.} $C= \Gamma_0$.  This action is $U({\cal N})$ invariant, although only the $SO({\cal N})$ sugroup is obvious in our Majorana spinor notation. 
The action is also super-Poincar\'e  invariant. The Poincar\'e Noether charges are
\begin{equation}
{\cal P}_m = P_m \, , \qquad {\cal J}^{mn} = 2 X^{[m} P^{n]} + \frac{i}{2} \bar\Theta_i \Pslash \Gamma^{mn}\Theta_i\, .
\end{equation}
The supersymmetry Noether charges are
\begin{equation}
{\cal Q}_i= \sqrt{2}\Pslash \Theta_i\, . 
\end{equation}

The massless superparticle action is also invariant under the following infinitesimal gauge transformations with commuting scalar parameter $\alpha(t)$ and 
anticommuting Majorana spinor parameters $\kappa_i(t)$
\cite{Siegel:1983hh} 
\begin{equation}
\delta X^m = \alpha P^m  + i \bar\Theta_i \Gamma^m \Pslash \kappa_i\, , \qquad \delta \Theta_i = \Pslash \kappa_i\, , \qquad 
\delta e = \dot\alpha +4i\bar\kappa_i \dot\Theta_i \, .
\end{equation}
Notice that the ``$\kappa$-symmetry'' gauge invariance is not associated with a constraint in the action (\ref{CBS}); a corollary of this is that the ``geometric'' term in this action defines
a closed 2-form that is {\it not } symplectic  because it is not invertible on the mass-shell.  Thus, one cannot read off canonical Poisson brackets from this action, and this complicates
the determination of the Poisson-bracket algebra of the Noether charges. There are various ways around this problem. One is to pass to the supertwistor form of the action, as we shall
do in the following section.  Another is to gauge fix; in Appendix A we show how light-cone gauge fixing leads to the conclusion, for ${\cal N}=1$,  that 
\begin{equation}\label{susyalg2}
\left\{{\cal Q}_\alpha,{\cal Q}_\beta\right\}_{PB} = -i \left(\Pslash \Gamma^0\right)_{\alpha\beta}\, . 
\end{equation}

\subsection{The super-Pauli-Lubanski pseudovector}

The Pauli-Lubanski  (PL) pseudo-vector is
\begin{equation}\label{PLvector}
L^m = \frac{1}{2} \varepsilon^{mnpq} {\cal J}_{np}{\cal P}_q \, , 
\end{equation}
where we use here the classical Noether charge realisation of the Poincar\'e charges. The special feature of this pseudovector is that it is translation invariant
in the sense that it has  zero Poisson brackets with the 4-momentum Noether charge ${\cal P}_m$. In fact, $L^m= hP^m$ for a massless particle, where
$h$ is the particle's helicity. 

The analogous construction for  the super-Poincar\'e algebra is surprisingly subtle. 
It can be shown  that the pseudovector \cite{Buchbinder:1998qv,Zumino:2004nb}
\begin{equation}
Z^m = L^m - \frac{i}{8} \bar{\cal Q}_i \gamma_5 \Gamma^m {\cal Q}_i
\end{equation}
has zero Poisson bracket with both ${\cal P}_m$ and ${\cal Q}_i$ provided that 
\begin{equation}\label{super-PLcons}
{\cal P}^2=0 \qquad \&  \qquad \calPslash {\cal Q}_i =0\, , 
\end{equation}
which imposes a limitation to massless particles. We have recently generalized this construction so that it also applies to 
massive particles \cite{Arvanitakis:2016wdn}, but the massless case will suffice here.  In this case one finds that 
$Z^m = H{\cal P}^m$ for the Poisson bracket realisation of the super-Poincar\'e algebra, where $H$ is a ``classical superhelicity''.
As we show in the Appendix, light-cone gauge fixing shows that $H=0$ for the superparticle action (\ref{CBS}). This could have been 
anticipated from the fact that $h=0$ for the usual spin-zero point particle action,  but  the interpretation in terms of superhelicity
requires consideration of the quantum theory, which we take up later. 

\subsection{Worldline CPT}

For our purposes it will be important that the superparticle action also has  a discrete PT symmetry that involves a worldline time reversal. The transformations are those of (\ref{CPT1}) 
supplemented by\footnote{Or $\Theta(t) \to i \gamma_5\Theta(-t)$,  but this is equivalent because $\Theta\to \gamma_5\Theta$ is an independent symmetry.}
\begin{equation}\label{CPTCPT}
\Theta_i(t) \to i \eta\, \Theta_i(-t)\,  \qquad (\eta=\pm1). 
\end{equation}
The necessity of the factor of $i$ is explained in section \ref{sec:prelim}.  As for the spinning particle, also  discussed in section  \ref{sec:prelim}, we allow for composition with the symmetry $\Theta_i \to -\Theta_i$ by introducing the sign $\eta$.  

It would be possible to choose  signs $\eta_i$, one for each of the ${\cal N}$ spinors $\Theta_i$, but  worldline CPT will not commute with $SO({\cal N})$ rotations unless the signs $\eta_i$ are all equal.
In the quantum theory, unequal $\eta_i$ leads to the existence of a $\bZ_2$ operator,  associated with a particular $SO({\cal N})$ rotation, that anticommutes with $K$, implying a doubling of the state space even if $K^2=1$. For this reason, no generality is lost in assuming that  $\eta_i=\eta$ for all values of the index $i$.

\section{Supertwistor formulation}

We now pass to the supertwistor form of the massless superparticle action \cite{Shirafuji:1983zd}. Here we follow the procedure  spelled out for Majorana spinor notation in \cite{Mezincescu:2013nta}.  
The first step is to solve the mass-shell constraint in terms of a commuting Majorana spinor $U$:
\begin{equation}\label{solveP}
P_m = \pm\,  \frac{1}{2} \bar U \Gamma_m U\, . 
\end{equation}
The top sign corresponds to a choice of positive energy ($P^0$) and the bottom sign to negative energy. In what follows we choose the top sign. 

Substitution for $P$ yields\footnote{We use the identity $(\bar U\Gamma U)\cdot\Gamma \equiv 2[U\bar U +\gamma_5U \bar U\gamma_5]$.}
\begin{equation}
\left(\dot X^m - i \bar\Theta_i \Gamma^m \Theta_i\right) P_m =  \bar W \dot U + i \left(\lambda_i \dot\lambda_i + \tilde\lambda_i \dot{\tilde\lambda}_i\right) + \frac{d}{dt} \left(\cdots\right)
\end{equation}
where 
\begin{equation}
\lambda_i = \bar U \Theta_i \, , \qquad \tilde\lambda_i= \bar U\gamma_5 \Theta_i\, , 
\end{equation}
and 
\begin{equation}\label{defW}
W= \Xslash U - i\Theta_i \lambda_i -i \gamma_5 \Theta_i \tilde\lambda_i\, . 
\end{equation}
This expression for $W$ implies the identity
\begin{equation}
\bar U \gamma_5 W  - \bar\xi_i \xi_i \equiv 0 \, , \qquad \xi_i= \lambda_i + i\tilde\lambda_i\, . 
\end{equation} 
In order to promote $W$ to an independent variable we must impose this identity as a constraint, using a Lagrange multiplier $s$.
This step yields the new action 
\begin{equation}\label{zeromass}
S= \int dt\left\{ \bar W \dot U  + i\bar\xi_i \dot\xi_i - 2s H\right\} \, , 
\end{equation}
where 
\begin{equation}\label{defsuperh}
H= h - \frac{1}{2} \bar\xi_i \xi_i\, , \qquad  h =  \frac{1}{2} \bar U \gamma_5 W\, . 
\end{equation}
We again have an action with one first class constraint, which generates a chiral $U(1)$ gauge invariance. The $U(1)$ transformations are 
\begin{eqnarray}\label{u1}
s(t) \to s- \dot\vartheta(t) \, ; && U(t) \to e^{-\vartheta(t) \gamma_5} U(t)\, , \qquad W(t)  \to e^{\vartheta(t) \gamma_5} W(t) \, , \nonumber \\
&&  \xi_i  \to e^{-i\vartheta(t)} \xi_i 
\end{eqnarray}
where $e^{i\vartheta(t)}$ is a map from the worldline to $U(1)$. Observe that\footnote{This requires the factor of $2$ in the Lagrange multiplier term of (\ref{zeromass}).}
\begin{equation}
s \to s  + i g^{-1}\dot g\, , \qquad g = e^{i\vartheta(t)}\, , 
\end{equation}
which shows that $s$ is a worldline $U(1)$ gauge potential. 

We now turn to the symmetries of the action (\ref{zeromass}).  Since $H$ is  an invariant  quadratic form of $SU(2,2|{\cal N})$, which is a cover of the ${\cal N}$-extended superconformal group, the action is manifestly superconformal invariant. 
In particular, it is super-Poincar\'e invariant with Poincar\'e Noether charges
\begin{equation}\label{twistPoinc}
{\cal P}_m = P_m \equiv \frac{1}{2}  \bar U\Gamma_m U \, , \qquad {\cal J}^{mn} = -\frac{1}{2}\bar U \Gamma^{mn} W \, . 
\end{equation}
Notice that there is no contribution to the Lorentz charge from the anticommuting variables because these are now Lorentz scalars. 
It is convenient to write the supersymmetry Noether charges as a sum of (complex) chiral and anti-chiral charges. To this end we first define
the (anti)chiral projections of $U$:
\begin{equation}\label{project}
U_\pm = \bP_\pm U \, , \qquad  \bP_\pm = \frac{1}{2}  \left(1\pm i\gamma_5\right)\, . 
\end{equation}
In this notation we have 
\begin{equation}\label{Qss}
{\cal Q}_i = {\cal Q}^i_+ + {\cal Q}^i_- \, , 
\end{equation}
where
\begin{equation}\label{Qss}
{\cal Q}^i_+ =  \sqrt{2} U_+ \bar\xi_i \, , \qquad {\cal Q}^i_- = \sqrt{2}  U_-  \xi_i \, . 
\end{equation}
Using the canonical Poisson bracket relations determined by the geometric part of the action (\ref{zeromass}), one finds that 
the only non-zero anticommutators of the (anti)chiral spinor Noether charges are
\begin{equation}\label{PBQs}
\left\{ {\cal Q}_\pm^i, {\cal Q}_\mp^j\right\}_{PB} = -i\delta^{ij}\bP_\pm \Pslash \Gamma^0\, . 
\end{equation}
This is equivalent, for ${\cal N}=1$, to (\ref{susyalg2}).

\subsection{Super-Pauli-Lubanski redux}

The (anti)chiral spinors $U_\pm$ and ${\cal Q}_\pm^i$ are equivalent to $2$-component complex Weyl spinors, but we prefer here
to use $4$-component spinor notation. The fact that $U_\pm$ and ${\cal Q}_\pm^i$ are complex does not prevent us from continuing 
to define their conjugates as Majorana conjugates. Thus 
\begin{equation}
\bar {\cal Q}_\pm \equiv  {\cal Q}_\pm ^T \Gamma_0 \quad \left( \Rightarrow \ \bar {\cal Q}_\pm P_\mp =0\right)\, . 
\end{equation}
Using this notation, the super-Pauli-Lubanwski pseudo vector $Z^m$ is 
\begin{equation}
Z^m = L^m + \frac{1}{8} \left(\bar{\cal Q}_- \Gamma^m {\cal Q}_+ - \bar{\cal Q}_+\Gamma^m {\cal Q}_-\right)
\end{equation}
The restrictions (\ref{super-PLcons}) that are needed for supertranslation invariance of this pseudovector are 
identically satisfied in the supertwistor formulation of the massless superparticle.

Using the expressions (\ref{Qss}) for  the (anti)chiral spinor Noether charges, we now find that 
\begin{equation}
Z^m = H {\cal P}^m\, , 
\end{equation}
where $H$ is the constraint function of the superparticle action (\ref{zeromass}).  This tells us  that $H$ is the  classical superhelicity,  and the superparticle constraint  tells us, at least formally,  that the superparticle has zero superhelicity, in agreement with the light-cone gauge result.  We say ``formally'' because $H$ includes a quadratic term in anticommuting variables. 
Strictly speaking,  the superhelicity interpretation applies only to the eigenvalues of the operator that replaces $H$ in the quantum theory.

\subsection{Worldline CPT redux}

The action (\ref{zeromass}) inherits from (\ref{CBS}) an invariance under the composition of worldline-time reversal with an internal $PT$ symmetry. 
In the supertwistor formulation of the massless superparticle, this discrete symmetry of the classical action acts via the transformations
\begin{eqnarray}\label{CPT3}
t\to -t\, ; &&\quad s(t)\to -s(-t) \quad W(t) \to  -W(-t)\, , \quad U(t) \to U(-t) \nonumber \\ 
&&\quad \xi_i(t) \to i\eta\, \xi_i(-t) \, , \quad  \bar\xi_i(t) \to i\eta\, \bar\xi_i(-t) \, . 
\end{eqnarray}
The relations (\ref{solveP}) and (\ref{defW}) are invariant provided that $X$ and $P$ transform as in (\ref{CPT2}), so this discrete transformation is indeed the 
supertwistor analog of (\ref{CPT2}).  Notice that these transformations imply that 
\begin{equation}\label{classflip}
H(t)\to -H(-t)\, . 
\end{equation}

\section{The quantum superparticle}\label{sec:section5}

In the quantum theory, and for standard operator ordering, 
\begin{equation}
\bar\xi_i\xi_i \to \frac{1}{2} \sum_{i=1}^{\cal N} \left[\hat\xi_i^\dagger, \hat\xi_i\right] = \sum_{i=1}^{\cal N} \hat \nu_i - \frac{{\cal N}}{2}\, , 
\end{equation}
where $\{\hat \nu_i; i=1,\dots,{\cal N}\}$ are  fermion number operators with eigenvalues $0,1$ and the $-\frac{1}{2}{\cal N}$ term is  sum of fermi oscillator zero point ``energies''.  
This suggests that the classical superhelicity $H$, defined  in  (\ref{defsuperh}) should be replaced by the operator
\begin{equation}\label{Hoperator}
\hat H = \hat h   - \frac{1}{2} \sum_{i=1}^{\cal N} \left[\hat\xi_i^\dagger, \hat\xi_i\right]  = \hat h - \frac{1}{2} \sum_{i=1}^{\cal N} \hat \nu_i + \frac{{\cal N}}{4}\, . 
\end{equation}

As a check, consider ${\cal N}=4$ with $h=-1$ when $\nu_1=\nu_2=\nu_3=\nu_4=0$. For this case the constraint $\hat H |\Psi\rangle=0$ yields a supermultiplet with helicities 
ranging from $h=-1$ to $h=1$, i.e. the ${\cal N}=4$
Maxwell supermultiplet.  As a further check, let us consider the implications for the quantum ${\cal N}=4$ superparticle of the discrete symmetry with transformations (\ref{CPT3}).
In the quantum theory this is realised on the canonical  operators  as
\begin{equation}
K \hat UK^{-1} = \hat U\, , \quad K\hat WK^{-1} = -\hat W \, , \qquad K\hat\xi_i K^{-1}  = \eta\, \hat\xi_i^\dagger \, . 
\end{equation}
This implies that 
\begin{equation}
K\hat h K^{-1} =-\hat h \, , \qquad K \left[\hat\xi_i^\dagger, \hat\xi_i\right] K^{-1} = - \left[\hat\xi_i^\dagger, \hat\xi_i\right] \, , 
\end{equation}
and hence
\begin{equation}
K \hat H K^{-1} = - \hat H \, , 
\end{equation}
as expected from the classical transformation (\ref{classflip}). Since $K$ realizes a space-and-time inversion and is anti-unitary it effects a CPT transformation. We therefore conclude that 
the ${\cal N}=4$ physical state condition $\hat H |\Psi\rangle=0$ preserves CPT, as expected because the  ${\cal N}=4$ Maxwell supermultiplet is CPT self-conjugate.

More generally, we will allow for the operator-ordering ambiguity inherent  in the definition of $\hat H$ by taking the
physical state condition to be 
\begin{equation}
\left(\hat H - \frac{c}{2}\right)|\Psi\rangle=0\, , 
\end{equation}
for some constant  $c$. The inclusion of this constant  has implications for CPT  invariance since 
\begin{equation}
K \left(\hat H - \frac{c}{2} \right) K^{-1} = -\left(\hat H - \frac{c}{2} \right)  - c\, , 
\end{equation}
which shows that CPT is broken when $c\ne0$. This is possible in a supertwistor formulation of the quantum theory because a supertwistor wave equation is not a local wave equation on  Minkowski spacetime.

We could allow for a non-zero constant $c$ already in the classical action by an addition to the action of the Worldline-Chern-Simons (WCS) term
$$
S_{WCS} = c\int \! dt \, s\, . 
$$
This is invariant under infinitesimal $U(1)$ gauge transformations, and hence under any finite $U(1)$ gauge transformation connected to the identity, but it is not invariant under ``large'' $U(1)$ 
gauge transformations.  However $\exp\{iS_{WCS}\}$ will still be invariant provided that $c$ is an integer.  To summarise, if the action (\ref{zeromass}) is $U(1)$ gauge invariant then we may choose $c$ to be an integer. For the ${\cal N}=0$ case, this means that we could choose $c=n$ to get a constraint  $h= n/2$. In other words, the helicity is half-integral, as expected\footnote{For a 3D particle the gauge group could be $\bR$ rather than $U(1)$, if the Lorentz group is taken to be the universal cover of $sl(2;\bR)$, and then the 3D spin is not quantised; see \cite{Klishevich:2001gy} for a different perspective on this issue.}.  
This implies a violation of CPT because the CPT conjugate state has helicity $-n/2$,  but CPT was already  broken ``classically'' by our choice of non-zero $c$.  Strictly speaking, it is not 
really a classical breaking of CPT because the coefficient of the WCS tern is $\hbar c$ when we re-instate factors of $\hbar$. 

What we have here is an example of a global $U(1)$ anomaly of the type analysed in \cite{Elitzur:1985xj}. 
Such anomalies have previously been  shown to have implications for the quantum mechanics of spinning particle models (with worldline supersymmetry) \cite{Howe:1989vn,Mezincescu:2015apa}.  Now we see that they are also relevant to  the quantum mechanics of superparticles.

\subsection{${\cal N}=1$ superparticle}

For ${\cal N}=1$ we have a single set of fermi-oscillator variables and hence a two-state system with states $|\pm\rangle$ such that 
\begin{equation}
\hat\xi |-\rangle =0 \, , \qquad \hat\xi^\dagger|+\rangle =0\, . 
\end{equation}
We also have a single fermi-number operator $\hat\nu = \xi^\dagger\xi$ with eigenvalues $\nu=0,1$. 
The superhelicity constraint for ${\cal N}=1$ tells us that the helicity is
\begin{equation}\label{N1hels}
h = \frac{1}{2} \nu- \frac{1}{4 } + \frac{c}{2} = \frac{c}{2} \pm \frac{1}{4}\, . 
\end{equation}
In this case, we cannot choose $c=0$ because that would violate the condition that $2h\in \bZ$. Instead, we must choose $c= n+\tfrac{1}{2}$ for some integer $n$, in which case we get  
the  superhelicity $n/2$ supermultiplet with helicities $n/2$ and $(n+1)/2$. 

Notice that the choice $c= \tfrac{1}{2} +n$ in (\ref{N1hels}) implies that 
\begin{equation}
\kappa \equiv h- \frac{\nu}{2} = \frac{n}{2}\, , 
\end{equation}
which tells us that $\kappa$ is the superhelicity. Given that $H$ is the eigenvalue of $\hat H$, and $\kappa$ the eigenvalue of $\hat h - \tfrac{1}{2}\hat\nu$, we see from the ${\cal N}=1$ case 
of the relation (\ref{Hoperator}) that 
\begin{equation}
H= \kappa + \frac{1}{4}\, . 
\end{equation}
In other words, for ${\cal N}=1$ we should have 
\begin{equation}\label{BK}
Z^m = \left(\kappa + \tfrac{1}{4}\right) P^m\, , 
\end{equation}
where $\kappa$ is the superhelicity.   This result can also  be deduced directly from the ${\cal N}=1$ supersymmetry algebra  \cite{Buchbinder:1998qv}. 

The main conclusion here is that quantization of the ${\cal N}=1$ superparticle yields an {\it irreducible} two-state supermultiplet of superhelicity $n/2$; as this is not CPT self-conjugate for any choice of 
the integer $n$, there is no symmetry principle that can fix $n$, although the simplest choice is obviously $n=0$. As our claim of irreducibility conflicts with claims made in \cite{Bergshoeff:1989uj} on the basis of calculations using the light-cone gauge, we show in the Appendix how light-cone quantization confirms our conclusions.

\subsection{Cancelling the CPT anomaly}

We explained above how the CPT anomaly of the ${\cal N}=1$ massless superparticle can be interpreted as arising from a quantum WCS term in the action (needed to cancel an anomaly in the chiral $U(1)$ gauge invariance).  This viewpoint also suggests a means of cancelling the CPT anomaly.  

Let $S_0$ denote the classical ${\cal N}=1$ superparticle action in supertwistor variables, 
\begin{equation}
S_0[U,W;\xi; s]  = \int\! dt \left\{\bar W \dot U + i\bar\xi\dot\xi - s \left(\bar U\gamma_5 W - \bar\xi\xi\right)\right\} \, , 
\end{equation}
and let $S_\pm$ denote this action after the addition of a WCS term with coefficient $\pm\tfrac{1}{2}$ (in units where $\hbar=1$): 
\begin{equation}
S_\pm[U,W;\xi; s] = S_0[U,W,\xi; s] \pm \frac{1}{2} \int \! dt \, s\, . 
\end{equation}
The choice of top sign for the WCS term corresponds to the choice $c= \tfrac{1}{2} +n$ for $n=0$ and the bottom sign corresponds to this  choice for  $n=-1$.  This means that 
the top-sign quantum superparticle has superhelicity $0$ while the bottom-sign quantum superparticle has superhelicity $-\tfrac{1}{2}$. These multiplets combined yield the 
reducible, but CPT self-conjugate, WZ supermultiplet, which suggests that we consider the combined action 
\begin{equation}\label{combined}
S[U,\tilde U,W,\tilde W, \xi,\tilde\xi; s,\tilde s] = S_+[U,W;\xi; s] + S_-[\tilde U,\tilde W;\tilde\xi; \tilde s]\, . 
\end{equation}
We now have a $U(1)\times U(1)$ gauge invariance, with gauge potentials $(s,\tilde s)$. The WCS terms with  coefficients proportional to $\hbar$ ensure that  this gauge
invariance is maintained in the quantum theory. Physical states are now tensor products states of the form 
\begin{equation}
|{\rm Phys}\rangle = |\Psi\rangle \otimes |\tilde\Psi\rangle\, , 
\end{equation}
subject to the physical state conditions 
\begin{equation}
\left[\hat H - \frac{\hbar}{2}\right]|\Psi\rangle = 0 \, , \qquad \left[\hat{\tilde H} + \frac{\hbar}{2}\right]|\tilde\Psi\rangle = 0\, . 
\end{equation}

The action (\ref{combined})  is invariant under the following transformation 
\begin{eqnarray}\label{CPT33}
t\to -t\, ; && s(t)\to - \tilde s(-t) \, , \qquad \tilde s(t) \to - s(-t) \, , \nonumber \\
&& W(t) \to  -\tilde W(-t)\, , \quad \tilde W(t) \to - W(-t)\, , \nonumber \\
&& U(t) \to \tilde U(-t) \, , \quad \tilde U(t) \to U(-t)\, ,  \nonumber \\
&& \xi(t) \to i\eta\, \tilde\xi(-t) \, , \quad \tilde\xi(t) \to i\eta\, \xi(-t) \, , \nonumber \\
&& \bar\xi(t) \to i\eta\, \bar{\tilde\xi}(-t) \, , \quad \bar{\tilde\xi}(t)  \to i \eta\, \bar\xi(-t)\, . 
\end{eqnarray}
This is a composition of the transformation (\ref{CPT3}) applied to each set of variables combined with an interchange of the two sets; neither is individually a symmetry
of the combined action $S=S_++S_-$, but the composition of them is. This discrete symmetry is realised in the quantum theory by an anti-unitary operator $K$ with the property that
\begin{eqnarray}
K\hat U K^{-1} &=& \hat{\tilde U} \, , \quad  K\hat W K^{-1} = - \hat{\tilde W}\, , \qquad  K \hat\xi K^{-1} =\eta\,  \hat{\tilde \xi}^\dagger\ \nonumber \\
K \hat{\tilde U} K^{-1} &=& \hat U  \, , \quad K \hat{\tilde W} K^{-1} = -\hat W\, ,  \qquad K\hat{\tilde\xi} K^{-1} =\eta\,  \hat{\xi}^\dagger\, . 
\end{eqnarray}
These conjugation relations imply that 
\begin{equation}
\label{K_for_WZ}
K \left[ \hat H- \tfrac{1}{2} \right] K^{-1} = - \left[\hat{\tilde H} + \tfrac{1}{2} \right] \, ,  \qquad  K \left[ \hat{\tilde H}+ \tfrac{1}{2} \right] K^{-1} = - \left[\hat H - \tfrac{1}{2} \right] \, . 
\end{equation}
The two separate physical state conditions are therefore exchanged, so the combined physical state condition is invariant, and therefore CPT is unbroken, as expected. 

From our discussion of the spinning particle in the Introduction, one might expect $K^2=-1$ from the fact that we now have two fermi oscillators. 
To investigate this, it is convenient to consider the operators
\begin{equation}
\Sigma_\pm = \frac{1}{\sqrt{2}} \left(\hat\xi \pm \hat{\tilde\xi}^\dagger\right)  \, .  
\end{equation} 
Together with their hermitian conjugates, these operators constitute a  new basis for the fermi oscillator variables since
\begin{equation}
\left\{ \Sigma_\pm ,\Sigma_\pm^\dagger \right\} = 1\, , \qquad \left\{\Sigma_+,\Sigma_-\right\}=0\, . 
\end{equation}
In this new  basis we have 
\begin{equation}
K \Sigma_\pm K^{-1} = \pm \Sigma_\pm\, .  
\end{equation}
An explicit  realization of the anticommutation relations in  terms of real $4\times 4$ matrices is 
\begin{equation}\label{Sigmarealize}
\Sigma_+  = \tfrac{1}{2}\sigma_+ \otimes \bI_2 \, , \qquad \Sigma_- = \sigma_3 \otimes \tfrac{1}{2}\sigma_+ \, . 
\end{equation}
Given that $K=K_0\circ K'$ for unitary operator $K'$, an explicit $4\times 4$ matrix realization of $K'$ is 
\begin{equation}
K' = \bI_2 \otimes \sigma_3\, . 
\end{equation}
This gives $K^2=1$, so we may impose $K|{\rm Phys}\rangle = |{\rm Phys}\rangle$ to get a CPT self-conjugate supermultiplet with just $4$ helicity 
states. This is the WZ supermultiplet. 

\subsection{${\cal N}=2$ and Kramers degeneracy }

For the ${\cal N}=2$ superparticle we define the ``hyperhelicity'' (${\cal N}=2$ superhelicity) operator to be 
\begin{equation}\label{Hyper}
\hat H = \hat h   - \frac{1}{2} \sum_{i=1}^2 \left[\hat\xi_i^\dagger, \hat\xi_i\right]  = \hat h - (\hat \nu_1 + \nu_2)  + \frac{1}{2}\, . 
\end{equation}
For the case in which  $h=-\tfrac{1}{2}$ when $\nu_1=\nu_2=0$,  the constraint $\hat H |\Psi\rangle=0$ yields, apparently,  a CPT self-conjugate supermultiplet with helicities 
$(-\tfrac{1}{2}, 0,0,\tfrac{1}{2})$, which is half a hypermultiplet. 

However, we now have two fermi-oscillator contributions to the constraint function. As for the ${\cal N}=4$ case discussed earlier, we have an anti-unitary operator $K$ in the quantum theory 
such that 
\begin{equation}\label{tohold}
K^{-1}\xi_i K^{-1}  = \eta\, \hat\xi_i^\dagger \, ,  \qquad K\hat\xi_i^\dagger K^{-1} = \eta\, \hat\xi_i \, . 
\end{equation}
As for ${\cal N}=4$, this implies that 
\begin{equation}
K \hat H K^{-1} = - \hat H \, , 
\end{equation}
so the physical state condition $\hat H|\Psi\rangle=0$ preserves CPT. 

So far, the analysis of the ${\cal N}=2$ case differs in no essential way from that of the ${\cal N}=4$ case. This is true for any even integer $\cal N$, in which case we have 
$\cal N$ fermi oscillator contributions to the superhelicity operator $\hat H$.  However, we saw in section \ref{sec:prelim}  that for ${\cal N}$ Fermi oscillators and an anti-unitary operator $K$ such 
that (\ref{tohold}) holds, then $K^2= \eta^{\cal N}(-1)^{\frac{{\cal N}({\cal N}-1)}{2}}$ For even ${\cal N}$ this simplifies to 
\begin{equation}\label{Kram}
K^2= (-1)^{\frac{{\cal N}}{2}}\,  \qquad \left(\tfrac{1}{2}{\cal N} \in \bZ^+\right). 
\end{equation}
Thus $K^2=1$ for ${\cal N}=4$ and we can impose the reality condition $K|\Psi\rangle =|\Psi\rangle$ on physical states; only then is it true that we have a supermultiplet of $2^{\cal N} = 8+8$ independent helicity states. For ${\cal N}=2$ we have $K^2=-1$ and hence $2\times 2^{\cal N} = 2(4+4)$ independent helicity states; these are the states of the hypermultiplet. 

The formula (\ref{Kram}) also implies that there is a Kramers degeneracy for the CPT self-dual ${\cal N}=6$ supermultiplet, which has maximum helicity $\tfrac{3}{2}$. This is indeed true since the scalars in this supermultiplet are in the ${\bf 20}$ of $SU(6)$, which is intrinsically complex.

\section{Discussion}

For massless particles in a four-dimensional Minkowski spacetime there are a number of peculiarities of the  representation theory of the ${\cal N}$-extended super-Poincar\'e algebra that are absent
in higher dimensions and for massive particles in four dimensions.  These include the fact that there is no {\it irreducible} CPT self-conjugate massless supermultiplet when ${\cal N}$ is odd, and that the ${\cal N}=2$ hypermultiplet 
is a doubled version of the apparently obvious candidate for a  CPT self-conjugate ${\cal N}=2$ supermultiplet.  We have explained these peculiarities by determing the fate in the quantum theory 
of a worldline time-reversing symmetry of the classical ${\cal N}$-extended superparticle action that we have called  ``worldline-CPT''. 

This classical superparticle symmetry turns out to be anomalous for odd ${\cal N}$, and this explains the absence of CPT self-conjugate massless supermultiplets for odd ${\cal N}$. In the supertwistor formulation  of the superparticle action this CPT anomaly is a close cousin of the parity anomaly of three-dimensional gauge theories with an odd number of Majorana spinor fields; 
it arises because of a quantum clash between the discrete symmetry and a $U(1)$ gauge invariance: when ${\cal N}$ is odd, gauge invariance requires the introduction of a 
CPT-violating Worldline-Chern-Simons (WCS) term with a half odd-integer coefficient (in units of $\hbar$). 

We have also shown how the CPT anomaly may be cancelled  by starting with a doubled superparticle action, and hence a $U(1)\times U(1)$ gauge invariance in the supertwistor formulation. If 
the $U(1)$ factors are associated with WCS terms having equal magnitude but  opposite sign coefficients then a composition of  worldline-CPT  with an interchange of the two sets of 
superparticle variables is an invariance of the action that does not suffer from a quantum anomaly. For ${\cal N}=1$, quantization then yields the CPT self-conjugate but {\it reducible}
massless Wess-Zumino  multiplet.

For even ${\cal N}$ there is no worldline-CPT anomaly and so one may quantise preserving CPT, which is realised by an anti-unitary operator  $K$, such that $K^2=\pm1$.
In this case, one gets a CPT self-conjugate supermultiplet, apparently of $2^{\cal N}$ states but  the correct number is actually $2\times 2^{\cal N}$ unless the superparticle wavefunction is a $K$-singlet; i.e. an 
eigenstate with eigenvalue $1$. This is possible if $K^2=1$ but not if $K^2=-1$. The doubling of states when $K^2=-1$ is a worldline example of Kramers degeneracy in systems that are time-reversal invariant. 
It turns out that $K^2=-1$ when $\tfrac{1}{2}{\cal N}$ is odd, which explains why the CPT self-conjugate ${\cal N}=2$ hypermultiplet has $8$ helicity states rather than $4$. 

There are various ways to see why the ${\cal N}=1$ massless superparticle yields, upon quantization, an irreducible supermultiplet of definite superhelicity (zero being the natural choice).
One way is to quantise in the light-cone gauge. Another is to first pass to the supertwistor formulation. In the first case, manifest Lorentz invariance is lost and in the second case manifest spacetime locality is lost.
As CPT invariance is a necessary feature of a spacetime-local Lorentz invariant wave-equation, it is hard to see how it could be possible to quantise the usual massless superparticle action in a way that preserves 
manifest Lorentz covariance. Any solution to this notoriously difficult problem would surely yield a CPT-invariant equation, which would necessarily imply a reducible supermultiplet, in contradiction to what one 
finds  in the light-cone gauge. The results described here appear to put this old problem in a new light, which may  have implications for the covariant quantization of the Green-Schwarz superstring.

\appendix
\section{Appendix: massless superparticle in light-cone gauge}

In this Appendix we present a quantization of the ${\cal N}=1$ massless superparticle in a light-cone gauge. First we 
define light-cone coordinates $X^m = (X^+,X^-,{\bf X})$, where ${\bf X} = (X^2,X^3)$ and 
\begin{equation}
X^\pm = \frac{1}{\sqrt{2}} \left(X^3\pm X^0\right) \, , \qquad P_\pm = \frac{1}{\sqrt{2}} \left(P_3\pm P_0\right) \, .  
\end{equation}
We then fix the time reparametrization and ``$\kappa$-symmetry'' gauge transformations by imposing the conditions
\begin{equation}
X^+(t) =t\, , \qquad \Gamma^+\Theta(t)=0\, , \qquad \Gamma^\pm = \frac{1}{\sqrt{2}} \left(\Gamma^3\pm \Gamma^0\right). 
\end{equation}
We will also use the following real representation for the Dirac matrices. 
\begin{eqnarray}
\Gamma^0 &=& \left(\begin{array}{cc} 0 & \sigma_3 \\ - \sigma_3 &0 \end{array}\right)\, , \qquad \Gamma^1= \left(\begin{array}{cc} \sigma_1 &0 \\ 0 & \sigma_1 \end{array}\right)\nonumber \\
\Gamma^2 &=& \left(\begin{array}{cc} \sigma_3 &0 \\ 0 & -\sigma_3 \end{array}\right)\, , \qquad \Gamma^3= \left(\begin{array}{cc} 0 & \sigma_3 \\ \sigma_3 &0 \end{array}\right)\, .
\end{eqnarray}
In this representation 
\begin{equation}
\Pslash = \left(\begin{array}{cc} \sigma_1P_1+\sigma_3P_2 & \sqrt{2}\sigma_3 P_+ \\ \sqrt{2}\sigma_3 P_- & \sigma_1P_1 -\sigma_3 P_2\end{array}\right)\, . 
\end{equation}
Also, 
\begin{equation}
\gamma_5 := \Gamma^0\Gamma^1\Gamma^2\Gamma^3 = \left(\begin{array}{cc} -i\sigma_2 & 0 \\ 0 & -i\sigma_2\end{array}\right)\, . 
\end{equation}

The gauge-fixing condition $\Gamma^+\Theta=0$ sets to zero the bottom two components of $\Theta$, which therefore takes the form 
\begin{equation}
\Theta = \frac{1}{\sqrt{P_-}} \left(\begin{array}{c} \varphi \\ 0 \end{array}\right) \qquad \varphi = \left( \begin{array}{c} \varphi_1 \\ \varphi_2\end{array} \right)
\end{equation}
for {\it two-component} real anticommuting spinor $\varphi$.  It is convenient to trade its components  for  the complex anticommuting variable
\begin{equation}\label{defmu}
\mu= 2^{\frac{1}{4}} \left(\varphi_1 + i\varphi_2\right)  \, , 
\end{equation}
in which case the light-cone gauge action is 
\begin{equation}\label{simpaction}
S= \int \! dt \left\{ \dot{\bf X}\cdot{\bf P} + \dot X^- P_- + i \bar\mu \dot \mu- \frac{|{\bf P}|^2}{2P_-}\right\} \, .   
\end{equation}
We have used here the fact that the Hamiltonian in the light-cone gauge is $P_+$, which we express in terms of ${\bf P}$ and $P_-$
by solving the mass-shell constraint.  This action is  still super-Poincar\'e invariant because the Noether charges are gauge invariant.  

Recall that the Lorentz Noether charges are 
\begin{equation}\label{Poincs}
{\cal J}^{mn} = 2 X^{[m} P^{n]} + S^{mn}\, ,\qquad S^{mn} = \frac{i}{2} \bar\Theta \Pslash \Gamma^{mn}\Theta\, . 
\end{equation}
The  non-zero components of $S^{mn}$ in light-cone gauge are
\begin{equation}\label{spins}
S^{12} =  \frac{1}{2} \bar\mu\mu\, , \qquad S^{-I} = \frac{1}{2P_-} \varepsilon^{IJ} P_J \left(\bar\mu\mu\right)\, .
\end{equation}
Recall too that that the supersymmetry charge can be written as 
\begin{equation}
{\cal Q} = {\cal Q}_+ + {\cal Q}_-\, , \qquad {\cal Q}_\pm = \bP_\pm {\cal Q} \, , 
\end{equation}
where $\bP_\pm$ is the chirality projection operator introduced earlier in  (\ref{project}). 
In light-cone gauge one finds that 
\begin{equation}\label{QtoS}
{\cal Q}_+ = 2^{-\frac{1}{4}} \left(\begin{array}{c} -\frac{i}{\sqrt{2}} {\cal S} \\ \frac{1}{\sqrt{2}} {\cal S}  \\ {\cal S}' \\  i {\cal S}' \end{array}\right) \, , \qquad 
{\cal S} = \frac{1}{\sqrt{P_-}} (P_1+iP_2)\mu\, , \quad {\cal S}' = \sqrt{P_- } \, \mu \, .
\end{equation}
Using the canonical Poisson bracket relations that follow from the action (\ref{simpaction}) we find, as expected, that
\begin{equation}\label{susyalg22}
\left\{{\cal Q}_\alpha,{\cal Q}_\beta\right\}_{PB} = -i \left(\Pslash \Gamma^0\right)_{\alpha\beta}\, . 
\end{equation}

In the light-cone gauge the constraints ${\cal P}^2=0$ and $\calPslash {\cal Q} =0$ are identities, as they were in the supertwistor formalism.
This tells us that the super-Pauli-Lubanski pseudovector $Z^m$ will take the form $HP^m$ for classical superhelicity $H$. A direct calculation
in the light-cone gauge yields $H=0$, confirming our earlier conclusion that  the classical superhelicity is zero.  A similar calculation of 
the PL pseudo-vector yields the result that 
\begin{equation}
L^m = h P^m \, , \qquad h= \frac{1}{2} \bar\mu\mu\, . 
\end{equation}
This is essentially the same result that we found in section \ref{sec:section5}  from the  $H=0$ constraint  of the supertwistor formulation, 
but with the complex anticommuting  variable $\mu$ replacing $\xi$. Moreover, the canonical anticommutation relations of $\mu$ are the same as
those of $\xi$, so the analysis of the spectrum of helicities proceeds, from this point on, exactly as before. In particular, 
we confirm our previous conclusion that the quantum ${\cal N}=1$ superparticle has two polarization states comprising an {\it irreducible} supermultiplet 
of definite superhelicity. 

Finally, we turn to the worldline CPT transformations of the light-cone gauge action (\ref{simpaction}). These follow directly from the 
transformations of (\ref{CPT1})  and (\ref{CPTCPT}) because they preserve the light-cone gauge-fixing conditions.
The result is
\begin{eqnarray}
t\to -t\, ; && {\bf X}(t) \to -{\bf X}(-t)\, , \quad X^-(t) \to - X^-(-t) \, , \nonumber \\
&& {\bf P}(t) \to {\bf P}(-t)\, , \qquad P_-(t) \to P_-(-t)\, , \nonumber \\
&& \mu(t) \to i\eta\, \mu(-t) \, , \qquad \bar\mu(t)  \to i \eta\, \bar\mu(-t)\, . 
\end{eqnarray}
for arbitrary sign $\eta$. 
Observe that $h(t)\to -h(-t)$, as before. And, again as before, this leads to the conclusion that worldline CPT is anomalous because
there is no irreducible CPT self-dual ${\cal N}=1$ supermultiplet. 

\section*{Acknowledgements} We that Eric Bergshoeff for helpful correspondence. 
A.S.A. and P.K.T.  acknowledge support from the UK Science and Technology Facilities Council (grant ST/L000385/1). 
A.S.A. also acknowledges support from Clare Hall College, Cambridge, and from the Cambridge Trust.


\providecommand{\href}[2]{#2}\begingroup\raggedright\endgroup

\end{document}